\documentclass[journal]{IEEEtran} 
\usepackage{booktabs} 
\usepackage{enumitem} 
\setlist[itemize]{leftmargin=*}

\usepackage{graphicx}
\usepackage[utf8]{inputenc}

\usepackage{amssymb}
\usepackage{calc}
\usepackage{amsmath} 
\usepackage{amsfonts}
\usepackage{mathtools}
\usepackage{color}

\newcommand{\leftq}{``}
\usepackage[caption=false]{subfig}

\begin{document}
\title{The Future is Unlicensed: Coexistence in the Unlicensed Spectrum for 5G}
\author{
	\IEEEauthorblockN{
	    Suzan Bayhan\IEEEauthorrefmark{1},  
	    G\"urkan G\"ur\IEEEauthorrefmark{2}
	    , and
		Anatolij Zubow\IEEEauthorrefmark{1}
		}\\
	\IEEEauthorblockA{
	\IEEEauthorrefmark{1}Technische Universit\"at Berlin, Germany,  e-mail: \{bayhan, zubow\}@tkn.tu-berlin.de}\\
	\IEEEauthorblockA{\IEEEauthorrefmark{2}Bogazici University, Istanbul, Turkey, e-mail: gurgurka@boun.edu.tr}\\
}
\maketitle




\begin{abstract}
	5G has to fulfil the requirements of ultra-dense, scalable, and customizable networks such as IoT while increasing spectrum and energy efficiency.
	Given the diversity of envisaged applications and scenarios, 
	one crucial property for 5G New Radio~(NR) is flexibility: flexible UL/DL allocation, bandwidths, or scalable transmission time interval, and most importantly operation at different frequency bands.
	In particular, 5G should exploit the spectral opportunities in the unlicensed spectrum for expanding network capacity when and where needed.
	However, unlicensed bands pose the challenge of ``coexisting networks", which mostly lack the means of communication for negotiation and coordination.
	This deficiency is further exacerbated by the heterogeneity, massive connectivity, and ubiquity of IoT systems and applications.
	Therefore, 5G needs to provide mechanisms to coexist and even converge in the unlicensed bands.
	In that regard, WiFi, as the most prominent wireless technology in the unlicensed bands, is both a key enabler for boosting 5G capacity and competitor of 5G cellular networks for the shared unlicensed spectrum.
	In this work, we describe spectrum sharing in 5G and present key coexistence solutions, mostly in the context of WiFi. 
	We also highlight the role of machine learning which is envisaged to be critical for reaching coexistence and convergence goals by providing the necessary intelligence and adaptation mechanisms.
\end{abstract}

\section{Introduction}\label{sec:introduction}
One ambitious goal set for 5th generation wireless networks (5G) is 1000x capacity increase for serving data-hungry and delay-sensitive applications in heterogeneous and ultra-dense network settings such as IoT and massive machine-to-machine communications.
While many approaches such as massive MIMO and device-to-device communications are being explored with this target in mind,  
5G vision is hard to realize with traditional cellular network architecture because of insufficient flexibility and scalability. In particular, capacity over-provisioning is needed to handle load dynamics. 
An operator deploys its network resources, e.g., equipment and spectral resources, considering the traffic demand forecasts as well as failure scenarios, to satisfy a certain level of performance. 
While spare capacity is needed to handle sudden increase in traffic load and is a safety measure against traffic uncertainty, the added idle capacity should be minimized to ensure cost-effectiveness. 
Note that macro-cell resources are typically over-provisioned by a factor of 5-10~\cite{rost2015benefits}.
The situation is even more severe with small-cells where traffic profile observed is less homogeneous and we see stronger traffic variation per cell, i.e., areas/hot-spots with peak traffic and areas with very low traffic, hence requiring an even higher over-provisioning of resources. 

To address these efficiency, scalability, and capacity issues, we believe that there is a need for disruptive solutions able to deliver
more capacity while not increasing the operational and capital expenditures. 
However, a critical bottleneck is the spectral inefficiency inherent to exclusive spectrum use, especially at precious and over-crowded sub-6\,GHz bands. 
Hence, spectrum sharing and especially operating in the unlicensed spectrum is vital for meeting 5G goals.
Specifically, unlicensed spectrum offers a myriad of opportunities to operators such as WiFi~(IEEE 802.11) offloading or multi-RAT operation.

\begin{figure*}
	\centering
	\includegraphics[width=0.6\linewidth]{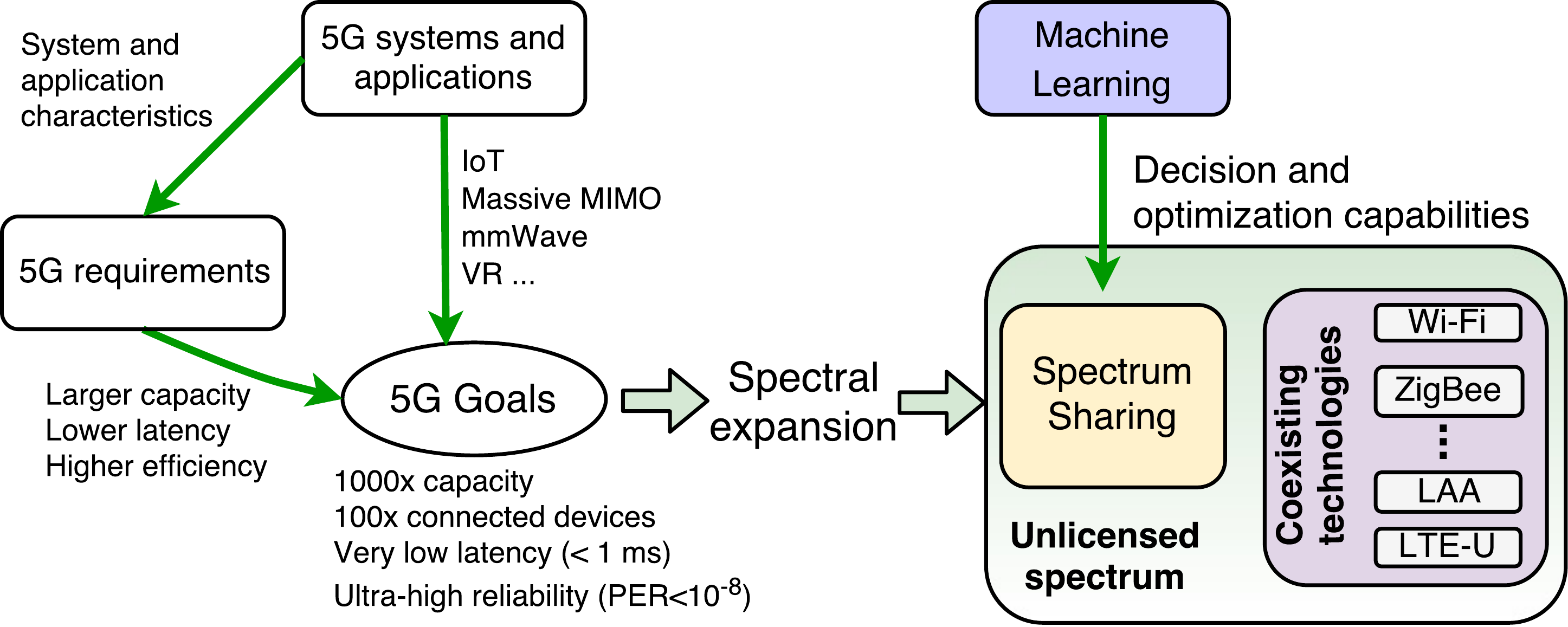}
	\caption{5G and spectrum sharing landscape. We focus on the right-hand side of this view, discussing how spectrum sharing in unlicensed bands powered by machine learning will realize robust spectral expansion and efficiency. 
	}
	\label{fig:overall}
\end{figure*}

\begin{figure}[ht]
	\centering
	\includegraphics[width=0.9\linewidth]{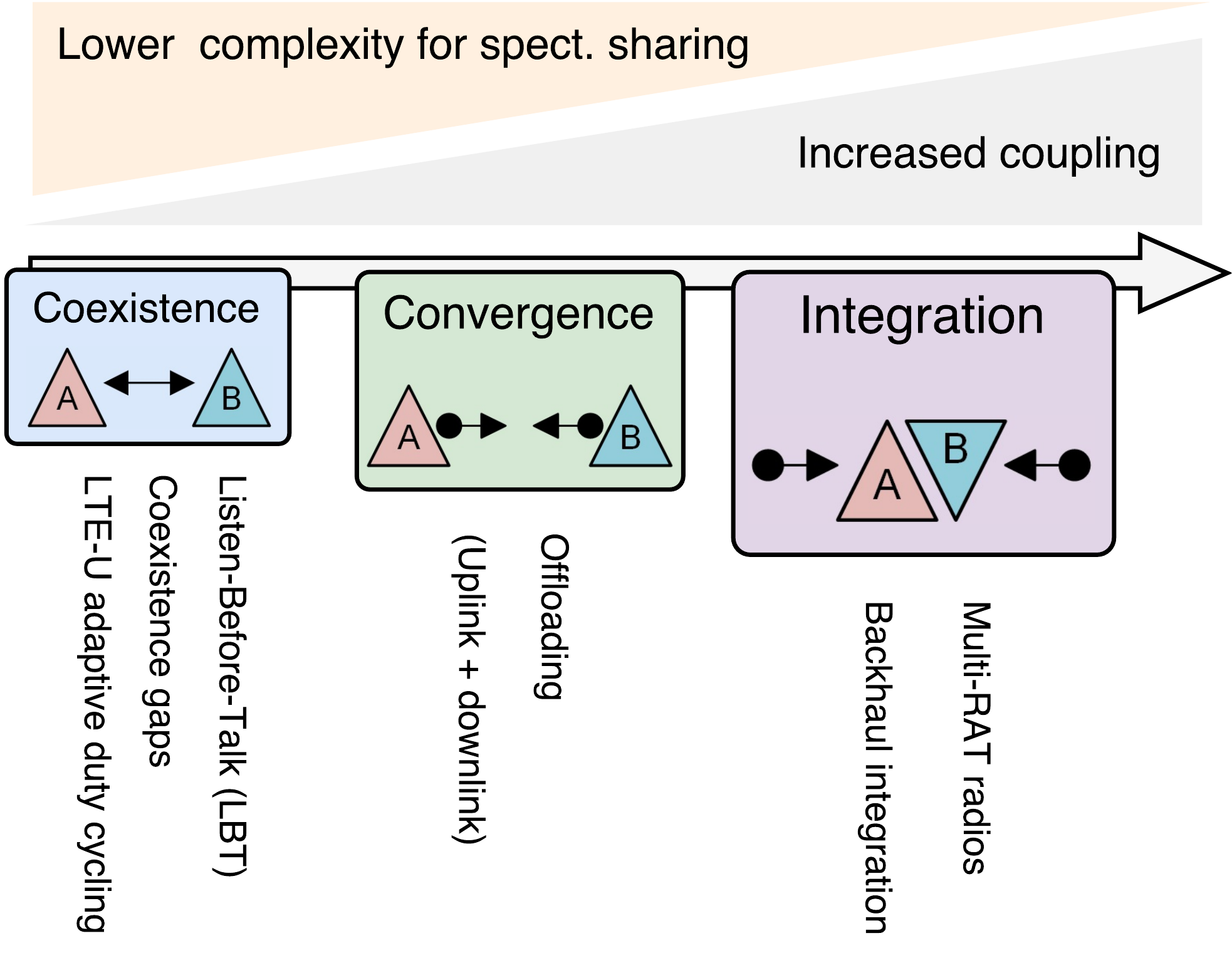}
	\caption{Levels of coupling between different radio access technologies for spectrum sharing. 
	}
	\label{fig:Levels}
\end{figure}

There are two key reasons why we reckon that unlicensed spectrum will continue to be a pillar of 5G: flexibility and cost-effectiveness.
First, the possibility to utilize unlicensed spectrum where and when needed adds flexibility and agility a mobile network operator with limited licensed-spectrum mostly seeks.
Instead of over-provisioning according to the peak load, now operators can expand their capacity using unlicensed spectrum in different ways, ranging from coexistence to integration modalities as shown in Fig.\ref{fig:Levels}.
Having such an elasticity in network capacity is essential also for massive IoT networks with diverse requirements and exhibiting typically high heterogeneity of traffic volume in time and space.
Also note that massive connectivity promised in 5G calls for spectrum authorization schemes which ensure that access to the spectrum is not a barrier for innovation and new players, e.g., due to high fees for spectrum access or slow approval process. 
To this end, a closely-related approach which was long ago presented by cognitive radio vision is to let radios access the spectrum dynamically by implementing agile and adaptive medium access. 
In contrast to strictly-regulated licensed access, dynamic spectrum access requires reliable coexistence mechanisms which ensure that coexisting systems do not jeopardize each other's operation. 
While light-licensing spectrum authorization~(e.g., licensed shared access) introduces certain level of flexibility in that regard, unlicensed spectrum access promises even more opportunities by relaxing command-and-control nature of license-based counterpart. 

The second driver for unlicensed spectrum utilization is primarily the cost aspect: with the pressing urgency of coping with the diminishing effective revenue per GB, operators are interested in the unlicensed spectrum which is spectrum-wise a free resource. 
Many regulatory bodies have also stated their viewpoint on the importance of unlicensed spectrum in unlocking the potential for 5G and thus opened new bands to unlicensed use, e.g., in 2016, FCC opened up 7 GHz spectrum  for unlicensed use out of 11 GHz spectrum opened in the mmWave bands. That expansion amounts to 7.5x of the existing unlicensed spectrum in the low frequencies~\cite{FCC_unlicensed2016}.

In a nutshell, given the 5G goals, applications, and their requirements depicted in Fig.\ref{fig:overall}, we believe that the future of wireless networking is unlicensed,  
which motivates us to investigate in this article the implications of operation in the unlicensed bands for 5G networks including IoT. 
As WiFi is the incumbent technology of the unlicensed spectrum, we also discuss how 5G networks can co-exist with, converge, or integrate to WiFi. 
Albeit its merits, unlicensed spectrum has its own limits, which we discuss later in the context of 5G. 

\section{Key Properties of 5G}\label{sec:5g}
5G is characterized by its heterogeneity of services and their requirements, e.g., from delay-tolerant applications running on resource-restricted devices like wearables to low-latency, high-reliability industrial applications running on high-capacity high-bandwidth devices. 
Meeting the requirements of these heterogeneous settings requires high flexibility of the 5G New Radio~(NR) and 5G network architecture.
The flexibility can be introduced in various ways~(e.g., time-domain or frequency domain~\cite{bhushan_5Gairint_2017}) and at various entities in the network~(e.g., end devices or the core network).
The radio itself needs to be agile and adaptive as rendered by the cognitive radio vision.
For example, the 5G NR air interface needs to be flexible to provide different channel bandwidths and carrier spacing, and to utilize non-continuous spectrum~\cite{bhushan_5Gairint_2017}. 
Moreover, the network itself can provide flexibility by new paradigms such as software-defined networking~(SDN) and Network Function Virtualization~(NFV).
This flexibility in the network architecture is primarily motivated by the cost-effectiveness of such logically centralized approaches.

IoT is envisaged to be an integral part of 5G ecosystem. Therefore, 5G is expected to support IoT devices for their diverse applications and service requirements. 
However, various challenges emerge regarding IoT and spectrum usage in 5G. 
First, IoT devices are expected to work together with cloud-resident services, which creates a much heavier uplink traffic compared to conventional cellular systems. 
Moreover, the stringent operational settings ranging from  ultra-reliable and low-latency scenarios to massive connectivity requires a much more flexible and agile radio. This difficulty is aggravated with the limited power and computational capabilities in many IoT devices \cite{IoTsparsity_commag17}. 
On the radio technology front, no one-fits-all solution exists as a result of heterogeneous IoT requirements such as in data rate, coverage region, and criticality. 
Hence, a plethora of wireless technologies are emerging for IoT support.
We can categorize them into two as proprietary~(e.g., Sigfox, LoRA) and open standards~(e.g., WiFi, ZigBee), or according to their operation spectrum as licensed~(e.g., LTE-MTC or NB-IoT, NB-LTE-M) and unlicensed spectrum technologies~(e.g., Sigfox, LoRA), or according to their coverage range as short-range or wide-range IoT. 
Current short-range IoT networks mostly operate in the unlicensed bands, whereas the wide-area IoT networks are expected to remain heavily reliant on cellular connectivity~\cite{spectrumForIoT2016}. However, for delay-tolerant applications, e.g., smart meters, providers can deploy wide-area IoT services in unlicensed spectrum in sub 1-GHz bands, e.g., TV white spaces, along with appropriate coexistence mechanisms.

\section{The role of WiFi in 5G} \label{sec:wifi}
As WiFi is an essential component in today's networking and cellular networks have benefited from WiFi in terms of mobile data offloading, 
3GPP has defined LTE-WLAN radio level inter-working in 3GPP Release 12~\cite{zhang2017ltesurvey}.
Moreover, Release 13 presents how LTE can expand to unlicensed bands at 5GHz by License Assisted Access~(LAA) where it has to coexist with incumbent WiFi networks. 
Additionally, LTE/WiFi aggregation~(LWA) defines how an LTE operator by a traffic split function at the eNB can use existing carrier WiFi infrastructure for traffic steering to WiFi networks.
Although it is hard to tell what exactly WiFi's role will be in 5G, given these developments, we can argue that it will remain equally important, if not more. 
Moreover, it is expected that WiFi offloading will still account for a significant share of mobile data traffic, e.g., almost fifty percent~\cite{cisco0217}. 
Given that a significant fraction of mobile data traffic is generated by indoor and stationary/nomadic users, WiFi, which is originally designed for delivering high capacity to stationary indoor users, can satisfy the service-level requirements of the cellular networks. 
The apparent proliferation and success of WiFi for wireless network access supports this expectation. 
For certain short-range IoT applications~(e.g., smart homes), we envision both unlicensed spectrum and WiFi as a specific technology to be a promising option.
To fully benefit from abundant WiFi infrastructure for 5G, efficient and practical coexistence~(as in LTE-Unlicensed) and even cooperation~(as in WiFi offloading) or integration schemes~(e.g., multi-radio base stations) are crucial.

\section{Spectrum Sharing in Unlicensed Bands} \label{sec:specsharing}
A coexistence scenario consists of at least two wireless networks or users who are in close geographical proximity such that the operation of one network can impair the operation of another markedly if they access the same spectrum band. But, they do not jeopardize performance of each other markedly thanks to the applied coexistence schemes. 
\subsection{Why is coexistence challenging?}
Coexistence among networks is challenging especially when these networks are heterogeneous in terms of underlying technology or run by different operators.
For the former, heterogeneity may imply the lack of common spectrum etiquette: the spectrum access rules differ across networks, which may hinder fair sharing of the resources.
A coexistence scenario consisting of an LTE unlicensed network~(LTE-U) and WiFi network is a representative example showing how difference in spectrum access etiquette may become troublesome~\cite{zhang2017ltesurvey}.
In this case, conventional LTE network does not implement listen-before-talk~(LBT) whereas WiFi does.  
As a result, WiFi nodes always defer the medium to LTE BS and therefore they may capture the medium only when LTE network does not have any traffic to transmit. 
The existence of a co-channel scheduled access network then leads to starvation and unfairness in spectrum sharing at the WiFi, if the former does not implement an efficient coexistence scheme.  
Hence, LTE-U applies duty-cycling, i.e., having off-periods during which WiFi can access the medium, and for the sake of fairness it adapts its off-period duration according to the observed WiFi traffic activity in the shared channel~\cite{bayhan_LTEnulling2017}. 

Moreover, heterogeneity might involve asymmetry in the coexisting networks, e.g., one network having a higher permitted transmission power level than the other such as in WiFi/ZigBee scenario. 
Even when regulatory bodies enforce that the networks accessing the unlicensed spectrum are restricted with a similar maximum transmission power level, e.g., 24 dBm for indoor LTE-U and WiFi\footnote{LTE-U Forum, http://www.lteuforum.org/uploads/3/5/6/8/3568127/lte-u\_forum\_lte-u\_technical\_report\_v1.0.pdf},
the increasing concerns on energy-efficiency especially for battery-powered cheap IoT sensors result in lower operation power for such devices.
Resulting power asymmetry  puts the low-power network in disadvantage and might result in strong interference from the other network in the lack of coexistence mechanisms.
In case of different operators, coexistence becomes challenging due to the competition among the operators or simply lack of interfaces for implementing collaboration among networks~\cite{zhou_coextoconverge_oct2017}.

\subsection{Mechanisms to improve coexistence capability of a network}

\begin{figure*}
	\centering
	\subfloat[Coexistence gaps in one dimension~(time-domain gaps, on the left) and coexistence in multiple dimensions~(time and space domain gaps, on the right).]{\includegraphics[width=0.50\textwidth]{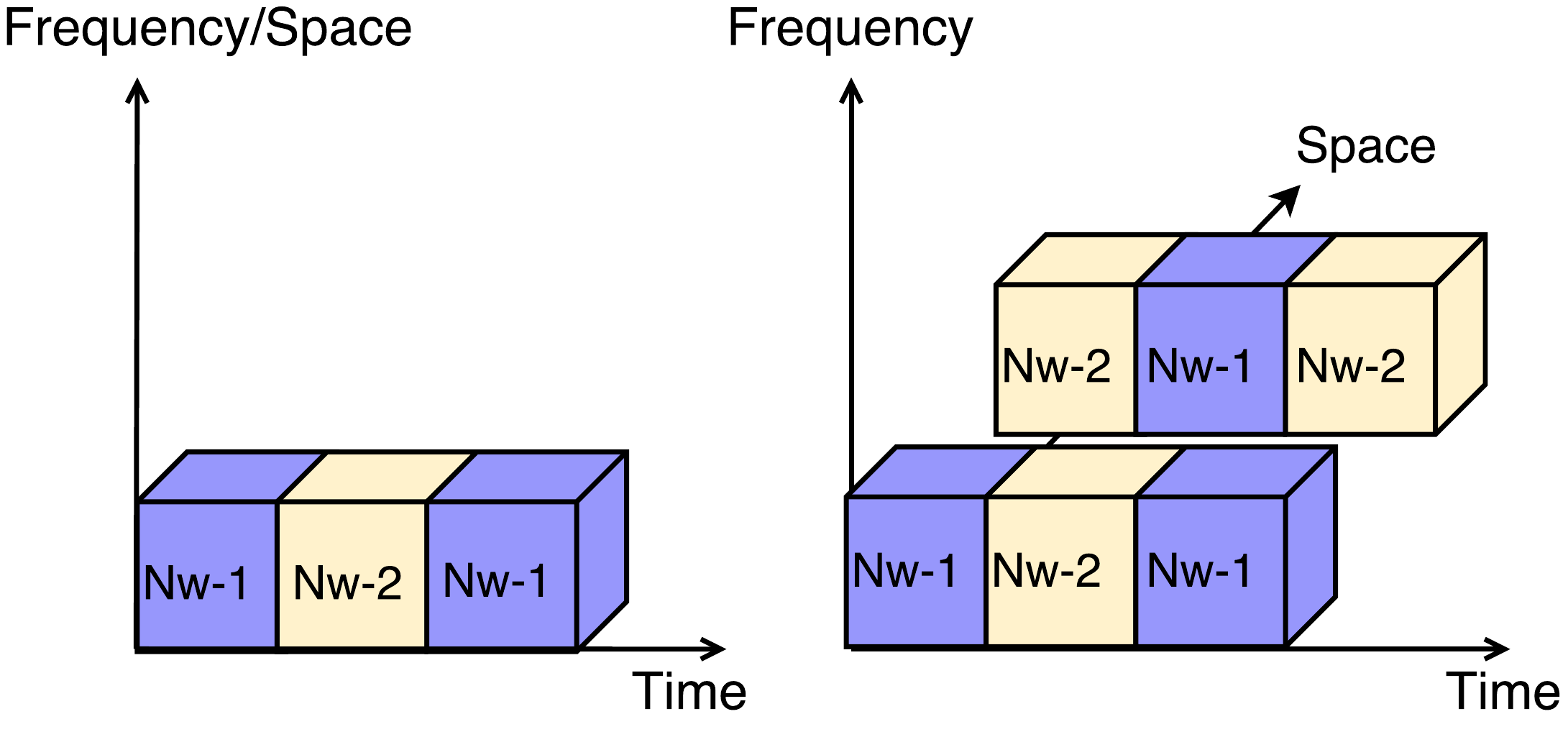}\label{fig:coexgaps}} \hfill
	\subfloat[Pareto front for the utilities of the two coexisting networks.]{\includegraphics[width=0.42\textwidth]{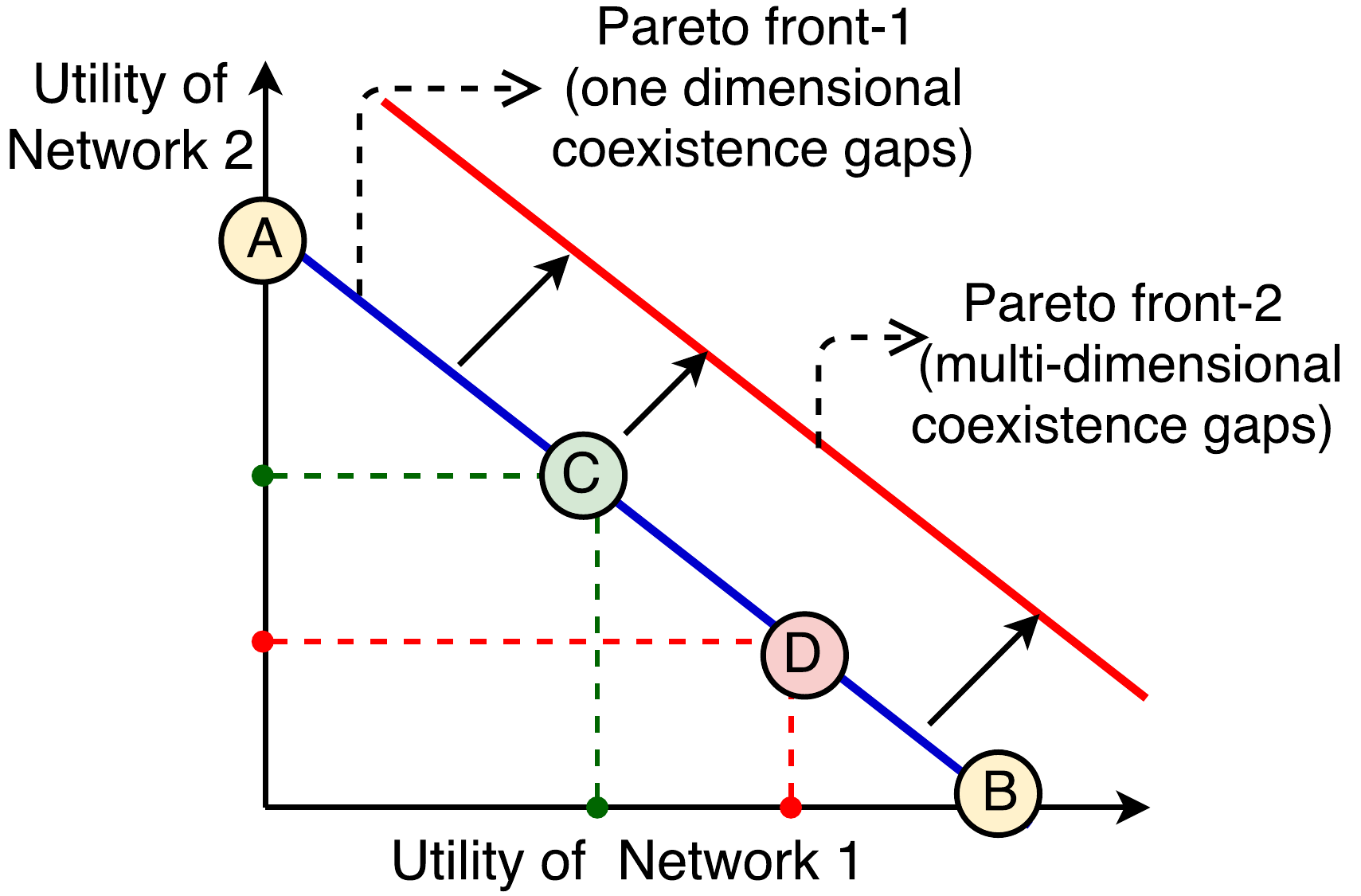}\label{fig:pareto}}%
	\caption{Coexistence gaps in one and multiple dimensions, and impact of coexistence gaps in achievable utilities.\label{fig:multidimensionalgaps}}
\end{figure*}

To ensure a peaceful coexistence, co-located systems or at least one of them introduces \textit{coexistence gaps}. 
Coexistence gaps can be in several domains: time, frequency, space, and code~\cite{bayhan_LTEnulling2017}.
Commonly, frequency domain gaps are implemented: networks sense the spectrum and select a clear channel 
which hosts no other network. So, colocated networks are separated in operation frequency.
LTE-U~\cite{zhang2017ltesurvey} 
implements frequency-domain gaps as the first step of its coexistence.
But, since spectrum is overly-crowded in dense urban areas, other coexistence gaps  are needed. 
In time domain, coexistence gaps correspond to the time periods when one network leaves the medium for others.
LTE-U puts time domain coexistence gaps by implementing duty-cycling. 
Coexistence gaps can be also put in the space domain by careful separation of the footprint of a network from the others. For example, cell-shrinking via power control is one way to implement coexistence gap in space domain. 
Similarly, a network can create \textit{almost blank spaces} by beamforming toward its receivers while applying interference nulling toward the users of other networks~\cite{bayhan_LTEnulling2017}. 
Given that the massive MIMO is a key component of 5G, space-domain gaps via interference nulling and beamforming can be used widely.
Finally, examples of code domain coexistence gap are CDMA or Resource Spread Multiple Access~(RSMA) defined in 5G for use cases requiring sporadic or grant-free  uplink transmissions
common in IoT.
Existing schemes create coexistence gaps usually in one of these dimensions.
However, multiple dimensions can also be exploited for maximizing the spectral efficiency as depicted in Fig.~\ref{fig:coexgaps}.

We argue that for successful coexistence, the networks should possess flexibility in at least one of the above-listed  dimensions.
For example, WiFi is coexistence-friendly as it exhibits agility in the time-domain, e.g., in the order of hundreds of $\mu$s, owing
to its LBT operation.
On the contrary, LTE lacks this merit due to its fixed-length frames and the nearest slot boundary is the earliest time 
it can react to coexistence-related challenges.
However, one can introduce flexibility in other dimensions to compensate for the inflexibility in one dimension.
For example, a network with time-domain inflexibility such as LTE-U can add coexistence gaps in the frequency or space domain, or both~\cite{bayhan_LTEnulling2017}.
A more ambitious option is to exploit all these dimensions and increase the coexistence capability of a network.
Note that a desirable property for a coexistence scheme is that it does not require substantial changes to the existing protocols and especially in the end user hardware. 
Hence, it becomes more challenging to exploit coexistence gaps in various dimensions by using already available functions and components of a technology.

Fig.~\ref{fig:pareto} shows the achievable utility region where Network-1 and Network-2 share the spectrum. 
Performance of the coexistence setting depends on the utilities of these two networks which could be for example throughput or energy-efficiency. 
In Point-A, Network-1 is using the whole resources resulting in highest utility for it. Similarly, Point-B is the case wherein Network-2 uses all resources alone. 
In fact, these points correspond to the cases only one network is active. 
When both networks are operational, each network could sustain only a lower utility than the corresponding upper bounds in Point-A and Point-B.
The first Pareto front in Fig.~\ref{fig:pareto} shows the points where one network's utility is maximized taking the other's parameters fixed. 
Whether Point-C or Point-D should be targeted or is achieved depends on the coexisting networks~(e.g., selfish or altruistic operation), coexistence scheme~(e.g., centralized or distributed), and the coexistence objective~(e.g., maximizing sum utility or individual utilities). 
By multi-dimensional coexistence gaps, networks can push the Pareto front to a more desirable one as in Fig.~\ref{fig:pareto}, e.g., achieving a higher performance for coexisting networks.

\subsection{Metrics for spectrum sharing efficiency and utilization} 
While the goal of coexistence schemes is to share the existing spectral resources in a way that all colocated networks can sustain certain level of performance, it is difficult to assess how performant a solution is. 
The common approach is to evaluate the throughput maintained by each network and the fairness of resource sharing.
However, as coexisting networks might be of different types, e.g., one cellular network while the other is an IoT network, traditional throughput-oriented analysis is inadequate for assessing coexistence schemes.
Hence, each network's performance goals should be considered in evaluation, e.g., for an industrial control network for fast event detection, guaranteed low latency is a primary performance goal whereas for a WiFi network, it is throughput. 
In WiFi/LTE-U coexistence, an LTE-U network is said to coexist well with a WiFi network if its impact on the WiFi performance is less adverse than another hypothetical WiFi network that would be deployed. 
In this definition of coexistence, WiFi is considered as the incumbent, similar to a primary network, and LTE-U network is expected to implement coexistence mechanisms. This is mostly due to the implicit agreement that WiFi plays a substantial role in current wireless communications and hence should be assured minimal impact from newcomers, i.e., cellular networks operating in the unlicensed bands.

Similarly, metrics for assessing the spectrum utilization efficiency are needed as the demand for radio spectrum increases so does the importance of its efficient use. 
While metrics like throughput measures technical efficiency of a system, e.g., the amount of data carried, such metrics lack notion of economic and functional efficiency. For example, the value generated by a technology might be more than another although its throughput capacity being lower than the latter.
Note that the economic value is more than the pure profit by a commercial entity or the license fees paid to the treasury but rather how much the use of the spectrum contributes to the society.
Regarding functional efficiency, it reflects functionality of a system involving hard-to-quantify subjective criteria such as ease of use~\cite{burns2002measuring}. 
Hence, new metrics capturing the diversity of technologies are essential. 
Finally, as not all bits are created equal, e.g., reliable delivery of public safety bits might be more crucial than the delivery of cellular traffic, coexistence and spectrum utilization metrics should  differentiate coexisting technologies by assigning different weights to each network.

\subsection{Coordinated vs. uncoordinated coexistence schemes} 
We can categorize spectrum sharing and coexistence scenarios into two as uncoordinated and coordinated schemes~\cite{aldulaimi_5gRace15}.
In the former, networks implement coexistence mechanisms on their own without any consultation from their neighbors whereas in the latter, networks directly or indirectly coordinate to ease their coexistence~\cite{Maglogiannis2017}. 
Majority of the existing solutions are uncoordinated as it does not require any infrastructure or change in the existing networks. 
On the other hand, coordinated solutions promise higher performance at the expense of higher complexity, e.g., \cite{ltfi_infocom18,bayhan_LTEnulling2017}.
The next section elaborates on these two schemes.

\section{Coordinated coexistence Schemes}\label{sec:coordinated}
In a coordinated coexistence setting, the coordinating networks have an infrastructure to exchange information about themselves or their expectations.  
Coordination can be implemented centrally, e.g., via SDN~\cite{sagari2015coordinated} or NFV~\cite{aldulaimi_5gRace15} techniques, or in a decentralized manner via a direct communication channel or in-band energy patterns if such a channel is nonexistent~\cite{Maglogiannis2017}. 
Moreover, coordination may require a common management/control plane
between heterogeneous technologies.
To exploit the opportunities of collaboration between heterogeneous wireless technologies, a cross-technology communication channel~(CTC), such as LtFi~\cite{ltfi_infocom18} is used.  
Finally, for efficient collaboration, a fine-grained
cross-technology proximity detection mechanism is needed~\cite{olbrichwiplus}. 
The communication channel can provide for example information about the identities of wireless network nodes in interference range and the level of mutual interference.

Coordinated solutions are expected to provide higher performance as coexisting networks can declare their requirements 
and operation parameters, in contrast to uncoordinated solutions where each network tries to identify first the existence of another network and then its operation parameters~\cite{Maglogiannis2017}. On the other hand, as there is no free lunch, this performance efficiency will come at the expense of infrastructure/protocol complexity and coordination overhead.

\subsection{Cross-technology Communication Channel for Coexistence Coordination}
A CTC can be used in many ways to facilitate the coexisting systems to make more informed decisions. 
Consider for example the coexistence between LTE-U/WiFi networks. A CTC control channel as in~\cite{ltfi_infocom18} can be used to perform cross-technology interference and radio resource management to avoid performance
degradation due to either contention, i.e., insufficient free airtime, or co-channel interference, i.e.,  packet corruption due to the insufficient sensitivity of the energy-based carrier sensing in WiFi (cross-technology hidden node). 
The former is achieved by adapting the LTE-U duty-cycling 
to the actual network load in both the LTE-U and WiFi networks to enable a fair use of the shared radio resources. 
Moreover, the WiFi MAC parameters like AIFS, CWmin/CWmax and TXOP can be adapted. Co-channel interference can be mitigated in two ways: first, by adapting the threshold used for energy-based carrier sensing in the WiFi network; second, by performing an interference-aware channel assignment to LTE-U and WiFi. 
Specifically, it is beneficial to put those networks~(LTE-U or WiFi) suffering from cross-technology hidden node problem on different unlicensed  channels. Cross-technology exposed terminal problem can be solved similarly.

\subsection{Where to deploy the coordination logic in the network?}
In a centralized coordination scheme, we need to determine the network location to deploy the coordination logic.
Considering the vast heterogeneity of 5G applications, one does not fit all: ultra-low latency industrial control applications call for fast coordination whereas delay-tolerant applications do not impose stringent constraints on 
the location of the coordination logic.
On the other hand, efficient coordination requires intelligence which might be hard to find in the network edge as a result of \textit{intelligent-core, dumb edge} trend~\cite{zhou_coextoconverge_oct2017}.
Note that operators favor centralized architectures due to their cost effectiveness. Additionally, 5G puts an emphasis on NFV and network slicing, resulting in low-end network edge.
For delay-tolerant applications, the required computation power at the edge can be populated by using the ample but distributed resources over the massive number of smart devices.
On the contrary, time-critical applications calls for stable resources at the edge. 
TV databases can be considered as an example of coordinated coexistence solution for white spaces.

In short, deciding on where to deploy the coordination logic is not straightforward due to the diversity of 5G applications. 
Rather than a single location, coordination functionality can be spread to different network segments, i.e., edge, fog, and the cloud. For example, frequency selection can  be coordinated by a cloud entity in a longer time scale whereas the network edge can help scheduling time-domain gaps in the selected frequency using a shorter time scale. 
In case of coordination at the edge, co-located networks can use CTC or wireless broadcast messages. 
For coordination in a wider coverage region beyond one-hop or two-hop neighborhood of the networks, communication takes place through the backhaul link.  
Depending on the distance between the networks and coordination entity, the communication overhead and coexistence efficiency will be different, usually with a trade-off between these two. 



\begin{figure*}[t]
	\centering
	\includegraphics[width=0.96\textwidth]{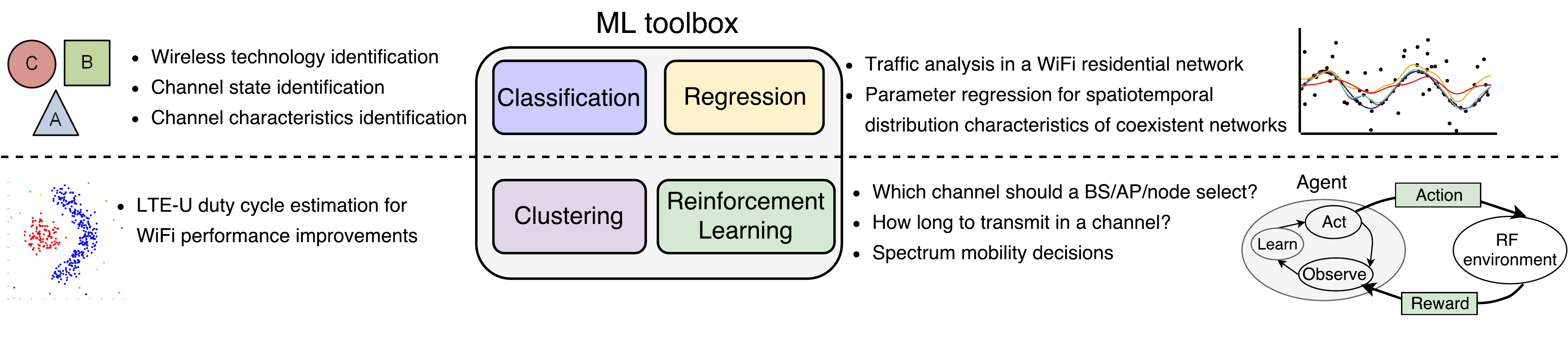} %
	\caption{Various spectrum sharing related questions and relevant ML functions. Please note that although some questions are stated in a technology-specific setting (e.g. LTE-U duty cycle estimation), they are also applicable to different wireless technologies, i.e., general challenges such as clear channel identification. 
	}
	\label{fig:CXZandML}
\end{figure*}

\section{Uncoordinated Coexistence Schemes}\label{sec:uncoordinated}

In an uncoordinated coexistence setting, the co-located networks  try to ensure coexistence without coordinating with the other networks and mostly based on their local observations. 
As we discussed in coexistence challenges, there is usually no mechanism for networks to coordinate, either due to heterogeneity of technologies or the diversity in the ownership of the networks. 
Therefore, most of the solutions fall into this category, e.g., an LTE-U network implements duty-cycling without communicating directly with the neighboring WiFi network(s).
While uncoordinated schemes have the merit of simplicity at a first glance, they might require more sophisticated techniques to implement neighbor-aware coexistence schemes.
For instance, an LTE-U network needs to decode the WiFi packets to estimate accurately the co-channel WiFi activity.
If an interface such as LtFi is available, WiFi activity and number of nodes can be directly obtained from the neighboring WiFi networks.
This would consequently decrease the overhead and complexity at the LTE-U nodes.

Despite the lack of coordination, networks can still develop techniques to share the spectrum efficiently by adapting to the dynamics of the coexistence setting through their observations of the wireless medium and their own performance.
For example, reinforcement learning can play an essential role in addressing the need for adaptive operation in a coexistence setting.
The next section discusses how machine learning~(ML) can help to address the challenges of coexistence and especially in case of uncoordinated schemes.

 
 \section{How can Machine Learning~(ML) help overcoming the challenges of coexistence?}\label{sec:ml}
 
 Real life scenarios are more heterogeneous and complicated than the current research addresses. For example, in contrast to simple settings of two co-located networks, there may be many different technologies from different operators in a practical environment.  
 Supporting such complex settings of 5G dense and heterogeneous deployments requires more than statically-defined rule-based schemes. 

 Moreover, to exploit coexistence gaps in many dimensions, we need context-aware solutions that can identify in which domains two systems can share the spectrum with high throughput efficiency. 
 Therefore, radios have to make some key decisions to achieve efficient and smart coexistence. 
 These decisions range from very fundamental ones such as identifying the occupancy of a channel to more advanced and complex ones such as traffic analysis for exploiting spatial WiFi characteristics.
 Considering these questions, ML provides a cognition toolbox to address the pertinent problems. Table~\ref{results} presents some key challenges and ML remedies. 
 For instance, ML enables identification schemes for recognizing coexistent communication systems and their inherent rules of operation. For the ``action" aspect, it yields adaptation-oriented intelligence, e.g., activation of LBT for WiFi coexistence or selection of spectrum mobility actions for specific coexistence contexts.
 
 \begin{table}[h]
 	\renewcommand{\arraystretch}{1.3}
 	\caption{How can ML function for spectrum sharing? 
 	}
 	\label{results}
 	\centering
 	\begin{tabular}{|p{2.2cm}|p{2.9cm}|p{2.3cm}|}
 		\hline
 		\textbf{Challenge} & \textbf{Phenomena} & \textbf{ML functions}\\
 		\hline \hline
 		Heterogeneity &  Different rules of operation, different ethics & Identification and adaptation\\
 		\hline
 		Power asymmetry & High-power systems vs. small-cell/low-power systems & Identification \\
 		\hline
 		Lack of communication among co-existing networks &  Networks controlled by different operators using different technologies & Detection and cooperation\\
 		\hline
 		Temporal and spatial network dynamics & Diverse characteristics (in time and space) of coexisting networks & Identification and adaptation\\
 		\hline
 	\end{tabular}
 \end{table} 
 
 Fig.~\ref{fig:CXZandML} shows a list of example questions regarding these challenges for spectrum sharing and ML functions. 
 Classification allows the radios to identify the type of neighboring networks~(e.g., an LTE or a radar network), channels and their characteristics~(addressing heterogeneity and lack of cross-technology communication) while reinforcement learning provides action-reward based decision support in unstructured environments. 
 However, these utilities also have some intrinsic challenges as listed below. 
 We identify those following points as open research directions that should be pursued to incorporate ML in 5G networks:
 \begin{itemize}
 	\item \textbf{Feasibility in practical settings}: Although ML-based capabilities are promising for spectrum sharing and coexistence, the practical viability of these schemes are not evident. 
 	\begin{itemize}
 		\item \textbf{Complexity}: Low complexity is necessary for cost-efficient and practical systems.
 		\item \textbf{Convergence time}: The ML-facilitated decisions should be timely to provide tangible benefits for system operation.
 		\item \textbf{What happens till convergence?}: The behavior of the coexistence scheme till the convergence is paramount since that may correspond to a significant portion of system time.
 		\item \textbf{Is it really possible to learn?}: Mobile or dynamic environments with transient behavior causing a \leftq cat-and-mouse" game brings forth learning challenges.
 	\end{itemize}
 	\item \textbf{Where to implement ML?}: The location of decision framework and its architecture~(centralized vs. distributed) render some intricate trade-offs such as latency vs. computational power or cost/practicality vs. robustness. 
 	Some candidate locations are wireless edge nodes, network-core, and cloud segments. 
 	The edge computing paradigm poses new opportunities to implement ML functions closer to the edge with collaboration from the cloud infrastructure. 
 	Considering IoT device diversity in terms of hardware capability and the application types, we envision that ML toolbox can be distributed in each of the above-listed components according to the processing needs as well as other performance metrics. 
 \end{itemize}

\section{Conclusions}\label{sec:conc}
Licensed cellular networks are limited in flexibility which is crucial for realizing 5G vision. 
In this article, we argue that unlicensed spectrum plays a key role for realizing the goals of 5G and beyond, by offering  cost-effective capacity expansion which is essential to address massive-scale and highly-diverse IoT networks. 
A challenge coming with unlicensed operation is coexistence among networks due to heterogeneity of co-located networks.
In this paper, we have provided some insights on coexistence solutions which is needed for spectrum sharing in 5G.
With network sharing and NFV becoming more prevalent in 5G, the coexistence challenges may be alleviated via centralized control and easier communication between different networks.
Finally, we envision that machine learning will play a fundamental role in making networks adapt to the coexistence setting for maximizing spectrum efficiency.

\bibliographystyle{IEEEtran}
\bibliography{biblio}
\end{document}